\documentclass[pra,preprint,superscriptaddress,floatfix,showpacs]{revtex4-1}

\usepackage{amsmath,amsfonts,amssymb}
\usepackage{graphicx}

\def\beq{\begin{equation}}
\def\eeq{\end{equation}}
\def\beqn{\begin{eqnarray}}
\def\eeqn{\end{eqnarray}}

\begin{document}

\title{Many-body tunneling dynamics of Bose-Einstein condensates and vortex states in two spatial dimensions}

\author{Raphael Beinke}
\email{raphael.beinke@pci.uni-heidelberg.de}
\affiliation{Theoretische Chemie, Physikalisch-Chemisches Institut, Universit\"at Heidelberg, Im Neuenheimer Feld 229, D-69120 Heidelberg, Germany}

\author{Shachar Klaiman}
\affiliation{Theoretische Chemie, Physikalisch-Chemisches Institut, Universit\"at Heidelberg, Im Neuenheimer Feld 229, D-69120 Heidelberg, Germany}

\author{Lorenz S. Cederbaum}
\affiliation{Theoretische Chemie, Physikalisch-Chemisches Institut, Universit\"at Heidelberg, Im Neuenheimer Feld 229, D-69120 Heidelberg, Germany}

\author{Alexej I. Streltsov}
\affiliation{Theoretische Chemie, Physikalisch-Chemisches Institut, Universit\"at Heidelberg, Im Neuenheimer Feld 229, D-69120 Heidelberg, Germany}

\author{Ofir E. Alon}
\affiliation{Department of Physics, University of Haifa at Oranim, Tivon 36006, Israel}

\date{\today}

\begin{abstract}
In this work, we study the out-of-equilibrium many-body tunneling dynamics of a Bose-Einstein condensate in 
a two-dimensional radial double well. 
We investigate the impact of interparticle repulsion and compare the influence 
of angular momentum on the many-body tunneling dynamics. 
Accurate many-body dynamics are obtained by solving the full many-body Schr\"odinger equation. 
We demonstrate that macroscopic vortex states of definite total 
angular momentum indeed tunnel and that, 
even in the regime of weak repulsions, a many-body treatment is necessary to capture the correct tunneling dynamics.
As a general rule, many-body effects set in at weaker interactions when the tunneling system carries angular momentum.
\end{abstract}


\maketitle

\section{Introduction}\label{sec:Intro}

The experimental realization of Bose-Einstein condensates (BECs) in trapped dilute ultracold gases \cite{anderson1995observation,Hulet1995,davis1995bose,RMP1,RMP2} 
paved the way for studying the dynamics of many-boson systems. 
In particular, tunneling phenomena with BECs have been of interest in recent years. 
A prominent example is the tunneling behavior in double-well potentials, 
often termed bosonic Josephson junctions in this respect \cite{gati2007bosonic}, 
which has been theoretically predicted \cite{milburn1997quantum,smerzi1997quantum} and observed in experiments \cite{albiez2005direct,levy2007ac}. 
Nowadays, the full many-body Schr\"odinger dynamics of a repulsive BEC in a one-dimensional 
double-well potential is available and shows the development of fragmentation, 
indicating that a many-body treatment is necessary in order to obtain the correct dynamics \cite{sakmann2009exact,PhysRevA.89.023602}.

Tunneling phenomena in two-dimensional (2D) trapped BECs, 
which is the main focus in this work, have been addressed recently. 
Especially the tunneling dynamics of trapped vortices were studied, e.g., 
in an harmonic potential with a Gaussian potential barrier \cite{martin2007transmission}, 
in 2D superfluids \cite{arovas2008quantum}, 
between two Gaussian wells \cite{salgueiro2009vortex}, 
or between two pinning potentials \cite{fialko2012quantum}.
In three-dimensional double-well potentials, 
macroscopic superpositions of vortex states during the tunneling dynamics 
have been found \cite{3DGarcia}. 

The purpose of this work is however to investigate the 
full many-body Schr\"odinger dynamics of a tunneling 2D BEC with definite total angular momentum.
To this end, we consider a 2D radial double-well trap (see Fig.~\ref{fig:Fig1}).
We discuss repulsive condensates made of $N=100$ bosons with zero total angular momentum, $L=0$, 
and vortex states with total angular momentum $L=N$. 
We demonstrate numerically that BECs carrying definite total angular momentum do indeed tunnel through the potential barrier. 
We compare the impact of angular momentum on the tunneling process and show that
many-body effects set in at weaker interactions when the tunneling system carries angular momentum. 
A general conclusion stemming from our results is that the long time tunneling dynamics 
of 2D BECs cannot be described by a standard mean-field, 
like the Gross-Pitaevskii (GP) equation, 
even in the regime of weak interaction between the bosons. 
Thus, one is in need of an accurate many-body theory. 
Our tool of choice is the multiconfigurational time-dependent Hartree for bosons (MCTDHB) method 
\cite{PhysRevLett.99.030402,alon2008multiconfigurational,book_MCTDH,book_nick}
which has been applied \cite{MCTDHB_OCT,MCTDHB_Shapiro,LC_NJP,Peter1,Peter2,klaiman2015variance} 
and benchmarked \cite{Benchmarks} 
in various numerical studies on repulsive many-boson systems in recent years, 
in particular for BECs in 2D traps, 
see Refs.~\cite{ALS,ALT,klaiman2014breaking,PhysRevA.91.063621,weiner2014angular,tsatsos2014vortex,klaiman2014spatially}.

The paper is structured as follows. 
In Sec.~\ref{sec:Theo}, we present the theoretical framework of our study, i.e., we describe the system's setup and introduce 
analysis quantities relevant for our purpose. 
The results are shown in Sec.~\ref{Results}, where we split the discussion of the
static considerations of the system (Sec.~\ref{static}) 
from the tunneling dynamics of $L=0$ (Sec.~\ref{dynL0}) and of vortex states (Sec.~\ref{dynvortex}). 
Concluding remarks are given in Sec.~\ref{Conclusion}. 
We provide additional information on the numerical preparation 
of vortex states at the many-body level as well as a brief discussion on the 
numerical convergence in appendices \ref{AppA} and \ref{AppB}.  

\section{Theoretical framework}\label{sec:Theo}

\subsection{Hamiltonian and setup}

The equation-of-motion for ultracold bosons in the 2D circular trap is the time-dependent Schr\"odinger equation
\beq\label{eq:SG}
 i\frac{\partial}{\partial t} |\Psi\rangle=\hat{H}|\Psi\rangle,
\eeq
where $|\Psi\rangle$ is the many-body wave-function which depends on the coordinates of all particles and time. 
The many-body Hamiltonian $\hat{H}$ is given by
\beq\label{Hamiltonian}
 \hat{H}=\sum_{i=1}^N \hat{h}(\vec{r}_i)+\sum_{i<j=1}^N \hat{W}(|\vec{r}_i-\vec{r}_j|),
\eeq
with the single-particle Hamiltonian $\hat{h}(\vec{r})=-\frac{1}{2}\Delta+\hat{V}(\vec{r})$, $\vec{r}=(r,\theta)$, comprised of the kinetic energy and the external potential $\hat{V}$, and the two-body interaction $\hat{W}$. 
We consider dimensionless units that are obtained by dividing $\hat{H}$ by $\frac{\hbar^2}{d^2m}$,
where $\hbar$ is Planck's constant, $d$ is a length scale, and $m$ the boson mass. 
For typical realistic parameters see Ref.~\cite{klaiman2014breaking}.

The short-range interaction between the bosons 
is modeled by a Gaussian function \cite{christensson2009effective,doganov2013two}, 
\beq
 \hat{W}(|\vec{r}_i-\vec{r}_j|)=\frac{\lambda_0}{2\pi\sigma}\,e^{-|\vec{r}_i-\vec{r}_j|^2/2\sigma^2},
\eeq
with $\sigma=0.25$. 
To quantify the interaction strength, we introduce the mean-field 
interaction parameter $\Lambda=\lambda_0(N-1)$ which is chosen 
to be positive in the following to describe repulsion between the particles. 

The external potential for the dynamics describes a 2D circular crater with a ring-shaped central
potential barrier, forming a 2D radial double well. 
It explicitly reads 
\beq
\hat{V}(\vec{r})=
\begin{cases}
	B\,e^{-2(r-R_B)^4}+C\,e^{-0.5\,(r-R_C)^4}     \,\,  &\mbox{if } r\leq R_C \\
 	C \,\, &\mbox{if } r>R_C
\end{cases},
\eeq
where $R_B$ and $R_C$ are the radial positions of the barrier and the crater's wall, and $B$ and $C$ their heights. 
Throughout this work, we set $B=1$ and the crater's wall is kept fixed at $R_C=9.0$ with constant height $C=200$. 
A schematic plot of the setup is shown in Fig.~\ref{fig:Fig1}. The chosen geometry serves as a natural way to combine tunneling phenomena in a double-well system with the conservation of angular momentum. Experimental realizations of this kind of trap have been achieved recently, see Refs. \cite{Corman,Mathew}.

The time-dependent Schr\"odinger equation Eq.~(\ref{eq:SG}) 
with the full many-body Hamiltonian Eq.~(\ref{Hamiltonian}) is solved by applying the MCTDHB(M) method. 
Its main idea is to express the wave-function of the system by a superposition of permanents $\{|\vec{n};t\rangle\}$
comprised of $M$ time-adaptive orbitals $\{\phi_j(\vec{r},t):1 \leq j \leq M\}$,
\beq
 |\Psi(t)\rangle=\sum_{\vec{n}}C_{\vec{n}}(t)\,|\vec{n};t\rangle,
\eeq
where $\vec{n}=(n_1,\ldots,n_M)$ is a vector carrying the individual occupation numbers of the orbitals and $\{C_{\vec{n}}(t)\}$ are the expansion coefficients. It is important to note that both the expansion coefficients and the basis set are time-adaptive and determined by the Dirac-Frenkel, time-dependent variational principle. 
For $M=1$, the MCTDHB theory boils down to the GP theory which is the standard mean-field commonly used 
to describe time-dependent BECs. 
Further details on the MCTDHB(M) method are given in the literature 
\cite{PhysRevLett.99.030402,alon2008multiconfigurational,book_MCTDH,book_nick}. 
We use the implementation in \cite{package}.
The simulations in this work are performed on 
a square box of size $[-12,12)\times [-12,12)$ with $128\times 128$ grid points. A translation into real units can be found in Ref. \cite{realunits}.
The obtained results are converged to the accuracy given below. 

\subsection{Quantities of interest}

In order to analyze the properties of the system, 
some useful quantities are needed to be introduced. 
The reduced one-particle density matrix of a many-body system is given by
\beqn\label{eq:RDM}
& & 
\rho^{(1)}(\vec{r}|\vec{r}^{\,\prime};t)=\langle\Psi|\hat{\Psi}^\dagger(\vec{r}^{\,\prime},t)\hat{\Psi}(\vec{r},t)|\Psi\rangle
= \sum_{j,k=1}^M \rho_{jk}(t)\,\phi_j^\ast(\vec{r}^{\,\prime},t)\phi_k(\vec{r},t)= \nonumber \\
& & \qquad \qquad = \sum_{k=1}^M n_k(t)\alpha_k^\ast(\vec{r}^{\,\prime},t)\alpha_k(\vec{r},t), \
\eeqn
with the annihilation operator $\hat{\Psi}(\vec{r},t)=\sum_j \hat{b}_j(t)\phi_j(\vec{r},t)$ that annihilates a boson at position $\vec{r}$ at time $t$. 
Its conjugate counterpart denotes the creation operator, creating a boson at position $\vec{r}$ at time $t$. 
The eigenvalues $\{n_k(t)\}$ of 
$\rho_{jk}(t)=\sum_{\vec{n}^\prime,\vec{n}} C^\ast_{\vec{n}^\prime}(t)C_{\vec{n}}(t)\langle \vec{n}^\prime;t|\hat{b}_j^\dagger (t) \hat{b}_k (t)|\vec{n};t\rangle$ 
are called the natural occupations and the eigenfunctions $\{\alpha_j(\vec{r},t)\}$ the natural orbitals of the reduced one-particle density matrix. 
The natural occupations are arranged in descending order, i.e., $n_1(t)\geq n_2(t)\geq\ldots\geq n_M(t)$, with $\sum_j^M n_j(t)=N$. 
If there is only one macroscopic eigenvalue, the system is said to be condensed \cite{penrose1956}, 
whereas the system is said to be fragmented if two or more eigenvalues are macroscopic \cite{nozieres1982particle,noz2006,spekkens,PhysRevA.78.023615,MCHB,muellerFragmentation,
baderFragmentation,fischerFragmentation,zhouFate,goldstoneFrag,songFrag,kangFrag,Yip}. 

\begin{figure}[!]
\includegraphics[angle=-90,width=0.8\columnwidth,angle=0]{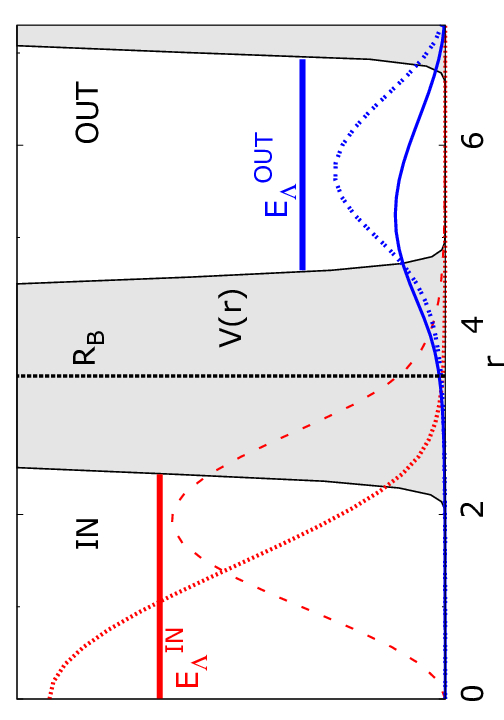}
\caption{(Color online) 
Schematic plot of the system's setup. 
Shown is a cut through the potential $V(r)$ (shaded gray area) 
of the 2D radial double well, 
as well as typical density cuts of the ground-state GP
orbitals $\phi_\Lambda^\text{IN}$ and $\phi_\Lambda^\text{OUT}$ for both $L=0$ and $L=N$. The number of bosons is $N=100$. The IN and OUT regions of the trap are separated by the potential barrier at $R_B$ (vertical dotted black line). 
For $L=0$, the density maximum in the IN region is located exactly in the trap's center, i.e., at $r=0$ (dotted red curve). 
On the other hand, for $L=N$, one can clearly see the node of $\phi_\Lambda^\text{IN}$ (dashed red). In the OUT region, $\phi_\Lambda^\text{OUT}$ for $L=N$ (blue, double dotted) is pushed further to the crater's wall than for $L=0$ (solid blue). Horizontal lines denote typical (angular-momentum-dependent) values of the ground-state energies at the mean-field level for the corresponding orbitals, i.e. $E_\Lambda^\text{IN}$ (solid red) for $\phi_\Lambda^\text{IN}$ and $E_\Lambda^\text{OUT}$ (solid blue) for $\phi_\Lambda^\text{OUT}$. 
See text for more details. 
All quantities are dimensionless.
}
\label{fig:Fig1}
\end{figure}

The density of the evolving system is given by the diagonal of Eq.~(\ref{eq:RDM}),  
\beq
 \rho(\vec{r};t)=\rho^{(1)}(\vec{r}\,|\vec{r};t).
\eeq
For the following studies on the tunneling dynamics, 
the trap is separated into an internal and an external part by 
setting the ring-shaped potential barrier at the position $R_B$. We refer to these subsystems as the 
IN and OUT regions (see Fig.~\ref{fig:Fig1}), 
representing the trap's center and the external rim, respectively. 
The occupation probabilities of the two parts are defined as 
\beq\label{P_IN}
 P_\text{IN}(t)=\frac{1}{N}\int_{r\leq R_B} \rho(\vec{r};t)d\vec{r}
\eeq
and
\beq\label{P_OUT} 
 P_\text{OUT}(t)=1-P_\text{IN}(t)=\frac{1}{N}\int_{R_B<r\leq R_C} \rho(\vec{r};t)d\vec{r},
\eeq
respectively (the density practically vanishes for $r>R_C$). 
We will among others use these probabilities to follow the dynamics in time.

\section{Results and Analysis}\label{Results}

\subsection{Static considerations}\label{static}

Before we start to study the dynamics, 
we want to analyze some static properties of the system which we will later use. 
Therefore, we first investigate the dependence of the ground-state energy 
on the position of the barrier $R_B$ for $N=100$ bosons at the GP, mean-field level ($M=1$). 
We use the same methodology as in Ref.~\cite{klaiman2014breaking} for $L=0$ and Ref.~\cite{klaiman2014spatially} for vortex states. The bosons are either trapped in the IN or OUT region separately. 
We would like to study the influence of the interaction strength $\Lambda$ and the total angular momentum $L$. We call the corresponding energies $E^\text{IN}_\Lambda$ and $E^\text{OUT}_\Lambda$ and denote the respective ground-state orbitals as $\phi^\text{IN}_\Lambda$ and $\phi^\text{OUT}_\Lambda$ (see Fig.~\ref{fig:Fig1} for more details). 
The respective ground states have been obtained by propagating in imaginary time, 
see Refs.~\cite{Benchmarks,MCHB} for further details.

\subsubsection{Impact of the interaction $\Lambda$}

The ground-state energies for different barrier locations $R_B$ are depicted in Fig.~\ref{fig:Fig2}a. 
The total angular momentum is zero, $L=0$. 
We consider the non-interacting single-particle case ($\Lambda=0$) and the weakly-interacting GP case ($\Lambda=2$). A discussion on the repulsion strengths in real units can be found in Ref. \cite{interaction}.
In both cases, there are certain radii where the ground-state energies for the IN and OUT regions coincide. 
These radii are termed the crossing points or simply the crossings in this work. 
They clearly depend on the repulsion strength. 
In the non-interacting case, we observe the crossing point at $R_1=3.271$, 
whereas for $\Lambda=2$ the crossing point is at a larger radius, $R_2=3.412$. 
These two radii are used in the dynamics below. 

We did the same simulation for ten times weaker interaction, $\Lambda=0.2$, 
and three times stronger interaction, $\Lambda=6$, 
and obtained, respectively, crossings at $R_B=3.287$ and $R_B=3.608$ (not shown). 
We conclude from this that an increase of the repulsion strength $\Lambda$ shifts the crossing point to a larger radius. 
From the viewpoint of trapping the particles in the IN region, 
this can be explained as follows. 
Due to the repulsion, the particles tend to separate from each other and the corresponding orbital is broadened. 
To account for the additional space needed in the trap's center, 
the barrier needs to be shifted further towards $R_C$, i.e., to a larger radius.

\subsubsection{Impact of the angular momentum $L$}
 
In the previous subsection, we have shown that increasing the repulsion strength leads to larger values for the crossing point. 
The results were obtained for zero total angular momentum ($L=0$). 
Here, we would like to study the impact of angular momentum $L$. 
Details on the numerical preparation of such vortex states as well as on the 
conservation and measurement 
of $L$ (during the time evolution) are given in Appendix \ref{AppA}.

Fig.~\ref{fig:Fig2}b shows the ground-state energies where the total angular momentum of the system is $L=N$. 
The crossing point is located at $R_3=4.089$ ($\Lambda=0$). 
This already shows that angular momentum strongly affects the position of the crossing point, 
comparatively stronger than the interaction strength. 
For a rough estimate, the crossing point for a relatively strong 
interaction $\Lambda=20$ with $L=0$ is at $R_B \approx 3.95$, 
which is still smaller than $R_3$. 
Another observation is the significant increase in energy compared to the system with $L=0$. 
The reason why angular momentum shifts the crossing point further towards the crater's wall can be explained by considering the centrifugal barrier originating from the kinetic energy operator in 2D. 
It acts like an additional repulsive potential, pushing the particles away from the trap's center. 
It also leads to the characteristic nodal structure of the respective 
GP orbital $\phi^\text{IN}_\Lambda$ (see dashed red curve in Fig.~\ref{fig:Fig1}).

\begin{figure}[!]
\includegraphics[angle=-90,width=0.6\textwidth]{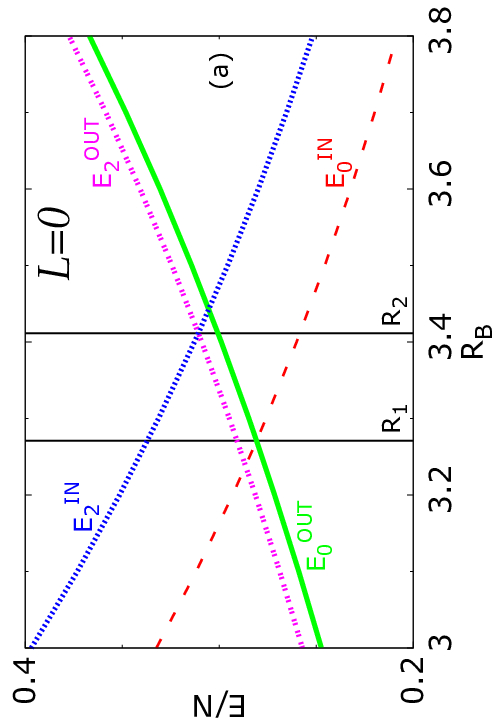}
\includegraphics[angle=-90,width=0.6\textwidth]{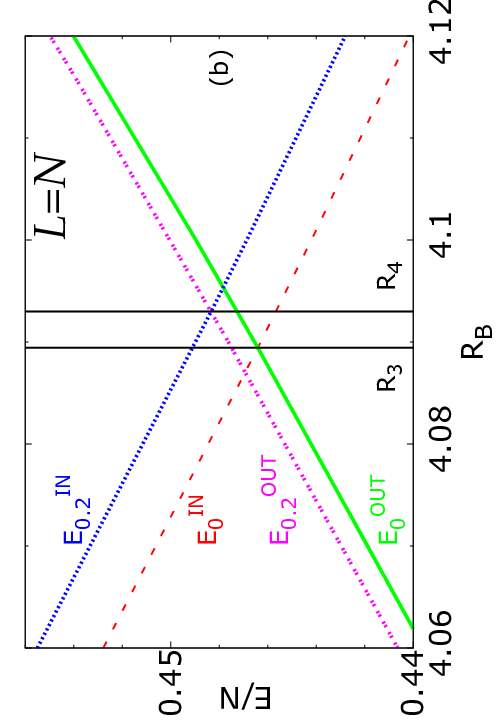}
\caption{(Color online) (a): Dependence of the ground-state energy on the position of the radial barrier $R_B$. 
The angular momentum is $L=0$, the number of bosons is $N=100$ and the number of orbitals used is $M=1$. $E_0^\text{IN}$ (dashed red curve) and $E_0^\text{OUT}$ (solid green) denote, respectively, the ground-state energies for the IN and OUT regions for $\Lambda=0$, $E_2^\text{IN}$ (dotted blue) and $E_2^\text{OUT}$ (magenta, double dotted) for $\Lambda=2$. 
The black vertical lines at $R_1=3.271$ and $R_2=3.412$ denote the crossing points for $\Lambda=0$ and $\Lambda=2$, respectively. Increasing the repulsion strength shifts the crossing point to larger radii. 
(b): Same as in (a) but for $L=N$. 
The corresponding crossing points are $R_3=4.089$ ($\Lambda=0$) and $R_4=4.093$ ($\Lambda=0.2$). 
The angular momentum $L$ affects the position of the crossing points comparatively stronger than the repulsion strength $\Lambda$. 
Note that the angular momentum is macroscopic; all particles carry it. 
See text for more details. 
All quantities are dimensionless.
}
\label{fig:Fig2}
\end{figure}

\subsection{Tunneling dynamics for $L=0$}\label{dynL0}

\subsubsection{The non-interacting system}

In this subsection, we analyze the tunneling dynamics for the non-interacting system ($\Lambda=0$), 
i.e., the single-particle case where the total angular momentum is zero, $L=0$. 
Thus, the orbital angular momentum is also zero, i.e., $l=0$ [see Eq.~(\ref{eq_A1}) in Appendix \ref{AppA} for more details]. 
Fig.~\ref{fig:Fig3}a depicts the time evolution of $P_\text{OUT}(t)$, Eq.~(\ref{P_OUT}), for different barrier positions $R_B$. 
The particles were prepared in the OUT region initially. 
In all cases considered, the bosons tunnel in a periodic manner between the two parts of the trap. 
The period and amplitude of the density oscillations clearly depend on $R_B$. 
Close to the crossing point $R_1=3.271$ (see Fig.~\ref{fig:Fig2}b), 
the amplitude is maximal; almost the whole density is involved in the tunneling process. 
Also the period of a single tunneling cycle is maximal. 
If one leaves $R_1$, both the amplitude and the period become smaller. 

Interestingly, the dynamics at the crossing point seem to be not sensitive to the initial condition, i.e., it does not matter whether the bosons are released from either the IN or the OUT region. 
This can be deduced from Fig.~\ref{fig:Fig3}b where the impact of the barrier location $R_B$ 
on the period $\tau$ and amplitude $A$ of the 
$P_\text{IN}(t)$- and $P_\text{OUT}(t)$-oscillations is shown. 
Whereas for $R_1$ there are no striking differences observable, 
the initial condition starts to matter as soon as the barrier location is varied. 
 
The period of the tunneling oscillations serves as a meaningful characteristic time scale for the dynamics. 
At the crossing point $R_1$, we obtain $\tau_1=\pi/J_1=110.231$ with 
$J_1=|\langle \phi^\text{OUT}_0|\hat{h}|\phi^\text{IN}_0 \rangle| =2.85\cdot 10^{-2}$. 
This is in a very good agreement with the result from Fig.~\ref{fig:Fig3}. 
$J_1$ has a similar physical meaning as the 
hopping parameter in the Bose-Hubbard model, see, e.g., Ref.~\cite{PhysRevA.89.023602}. 
It can be seen as a measure for the (unnormalized) transition probability amplitude
from the state $|\phi^\text{IN}_0 \rangle$ to the state $|\phi^\text{OUT}_0 \rangle$ under the influence of $\hat{h}$. 
Thus, the larger $J_1$ is, the faster a particle tunnels through the barrier. 
We will use the same time scale for the case of interacting bosons in the next subsection.

\begin{figure}[!]
\includegraphics[angle=-90,width=0.6\textwidth]{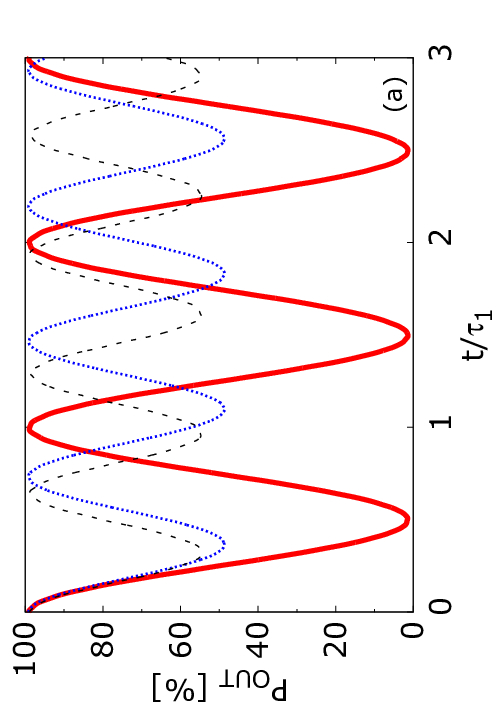}
\includegraphics[angle=-90,width=0.6\textwidth]{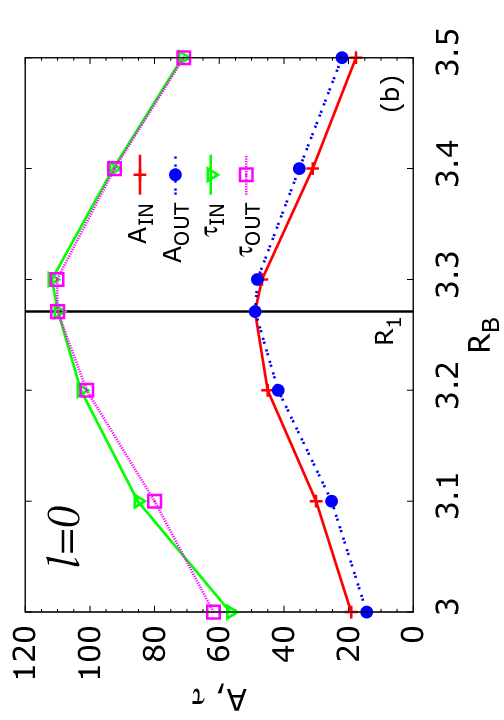}
\caption{(Color online) (a) Single-particle tunneling dynamics for different barrier positions $R_B$. 
The orbital angular momentum is $l=0$. 
The dynamics are started from the OUT region. 
Close to the crossing point $R_1=3.271$, the amplitude of $P_\text{OUT}(t)$ 
is maximal and almost the whole density tunnels (red solid curve). 
The period is in a very good agreement with the predicted value $\tau_1=110.231$. 
Leaving the crossing point, both the amplitude and period become smaller 
($R_B=3.1$, dotted blue; $R_B=3.5$, dashed black). 
(b) Dependence of the amplitude $A$ and period $\tau$ (plotted on the same axis) 
on $R_B$ for the same parameters as in (a). 
The indices IN and OUT denote the two different initial conditions considered. 
Close to the crossing point $R_1$, both the amplitude and period become maximal and give the same result for the two initial conditions. 
See text for more details. 
All quantities are dimensionless.
}
\label{fig:Fig3}
\end{figure} 
 
\subsubsection{The interacting system}

We now turn to the impact of the interparticle repulsion $\Lambda$ 
on the tunneling dynamics for $L=0$, both at the GP, mean-field 
level and especially at the many-body level. 
We split the discussion of the results to two interaction strengths, $\Lambda=2$ and $\Lambda=6$, 
which we term in what follows weak and strong, respectively. We point out that the distinction between `weak' and `strong' implies a classification with respect to the underlying physics. Whenever two modes faithfully describe the dynamics, we refer to the interaction as weak, whereas the interaction is termed strong whenever more modes are needed. Further physical distinctions between the dynamics with weak and strong interactions are discussed below. \\ 
It will turn out that for the many-body dynamics, numerical convergence 
for the regime of weak repulsion is reached with $M=4$ time-adaptive orbitals. 
We benchmark this result in Appendix \ref{AppB}. 
Due to the observations from the non-interacting system, 
we will focus on the tunneling dynamics at the corresponding crossing points. 
To recall, these radii depend on the interaction strength.

\begin{figure}[!]
\includegraphics[angle=-90,width=0.6\textwidth]{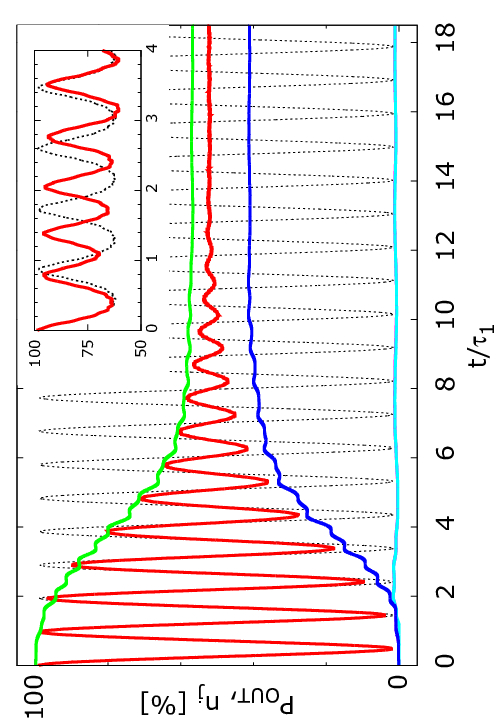}
\caption{(Color online) Mean-field ($M=1$) and many-body ($M=4$) tunneling dynamics of the interacting system at the crossing point $R_2=3.412$. The angular momentum is $L=0$, the number of particles is $N=100$, and the interaction strength is $\Lambda=2$. 
The bosons are initially prepared in the OUT region. 
Whereas the GP theory predicts unperturbed oscillations of $P_\text{OUT}(t)$ (dotted gray curve), 
the amplitude is damped at the many-body level (solid red) and saturates after approximately 14 cycles. 
Only the first two natural orbitals are significantly occupied (solid green and blue), 
the other two (solid magenta and light blue; curves lie atop of each other) carry just a small fraction of the particles. 
However, the frequencies of the GP and many-body density oscillations are essentially the same. 
Inset: Same simulation but for strong repulsion ($\Lambda=6$) at the respective crossing point $R_B=3.608$. 
Many fewer particles tunnel back and forth; the dynamics obtained from the GP and MCTDHB(4) descriptions start to deviate from each other already during the first cycle, both in their frequencies and amplitudes. 
After 4 cycles, the four natural orbitals are already macroscopically occupied (not shown). 
Note the different time scales. 
See text for more details. 
All quantities are dimensionless.
}
\label{fig:Fig4}
\end{figure}

To study the impact of weak repulsion, we set $\Lambda=2$ and put the barrier at $R_2=3.412$ (see Fig.~\ref{fig:Fig2}a). 
We obtain similar results for starting either from the IN or the OUT region and will therefore only concentrate on the latter. 
Fig.~\ref{fig:Fig4} shows the time evolution of $P_\text{OUT}(t)$ at the mean-field (GP) and many-body [MCTDHB(4)] levels. 
For the first cycle, the two approaches give very similar results. 
However, differences set in afterwards. 
On the one hand, the mean-field dynamics show essentially unperturbed oscillations which are very similar to the non-interacting case. 
Only the period is slightly shorter due to the repulsion. 
On the other hand, the oscillations in the many-body case show damping of the amplitude. 
$P_\text{OUT}(t)$ saturates after approximately 14 cycles with 52 particles in the external rim. 
This is very close to the equilibrium distribution of confining exactly 50 bosons in each part of the trap. 
Deviations from this can be explained with the accuracy of our numerical procedure to determine the crossing points. 
Thus, the observed behavior is similar to the density collapse in a bosonic Josephson junction, 
see, e.g., Refs.~\cite{gati2007bosonic,milburn1997quantum,PhysRevA.89.023602}. 
There, the density oscillations between the two wells are suppressed after some time.

The evolution of the natural occupations show that the many-body 
dynamics are dominated by two natural orbitals, i.e., the system evolves from 
being essentially fully condensed [$n_2(0)=\mathcal{O}(10^{-3})$, $n_3(0)$ and $n_4(0)$ are even smaller] to being two-fold fragmented. 
After the density oscillations have collapsed, the occupation numbers of the first two natural orbitals, $\alpha_1$ and $\alpha_2$, are approximately 56.7\% and 41.1\%, respectively. 
The remaining two orbitals are significantly less occupied (approximately $1\%$). 
It is interesting to note that the curves for $n_1(t)$ and $n_2(t)$ tightly envelope the oscillations of $P_\text{OUT}(t)$, 
indicating that fragmentation and the collapse of the density oscillations are closely related, see Fig.~\ref{fig:Fig4}. 
 
\begin{figure}[!]
\includegraphics[angle=0,width=0.6\textwidth]{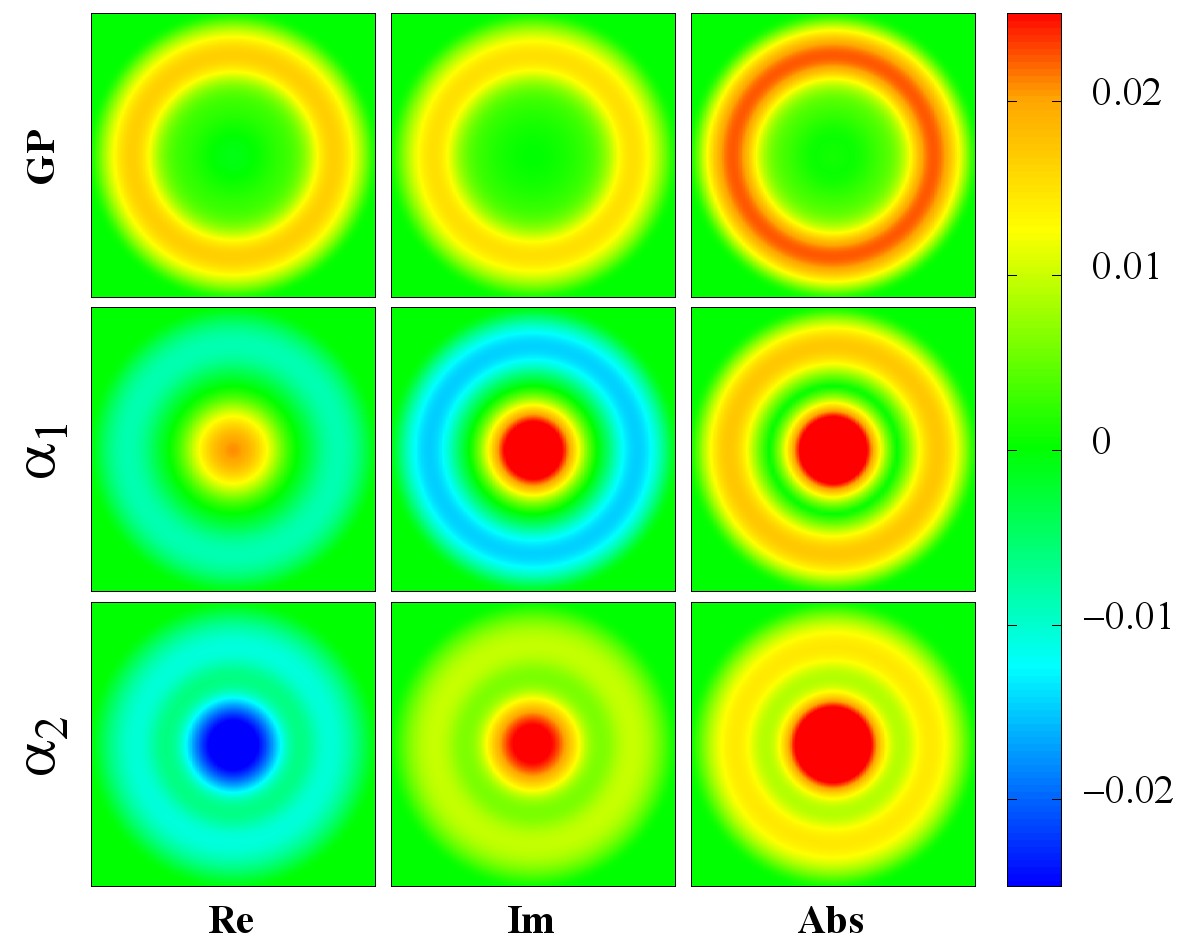}
\caption{(Color online) Real, imaginary, and absolute value of the GP, 
mean-field orbital (top panels) and of the first two natural orbitals $\alpha_1$ and $\alpha_2$ of the many-body simulation ($M=4$, middle and bottom panels) at $t= 18.4\,\tau_1$. 
The other parameters of the system are the same as in the main figure of Fig.~\ref{fig:Fig4}. 
Whereas the mean-field orbital is localized in the external rim, $\alpha_1$ and $\alpha_2$ from the many-body computation are delocalized, covering both the IN and OUT regions. 
See text for more details. 
All quantities are dimensionless.
}
\label{fig:Fig5}
\end{figure}

Let us have a closer look into the structure of the 
first two natural orbitals $\alpha_1$ and $\alpha_2$ at $t=18.4\,\tau_1$, i.e., 
when the density has already collapsed at the many-body level. 
The naive notion would be to have one localized orbital in the IN region and another one localized in the OUT region. 
However, Fig.~\ref{fig:Fig5} shows that they are rather delocalized in space, i.e., distributed over both subsystems. 
In contrast to that, the GP orbital is localized 
in the external rim and does not account for the occupation of the trap's center. 
This indicates a major difference between the fragmented many-body system 
and the condensed system when described at the GP level. 

The above simulations have been repeated for ten times weaker interaction, $\Lambda=0.2$, leading to qualitatively similar results. 
The only crucial difference at the many-body level is that it takes much longer (more than 200 cycles) 
until the density oscillations are completely suppressed. 

We now turn to the case of strong repulsion ($\Lambda=6$) with crossing point $R_B=3.608$. 
The occupation numbers at $t=0$ are: $n_1(0)=99.37$\%, $n_2(0)=n_3(0)=0.28$\% and $n_4(0)=0.06$\%, i.e., the system is still essentially fully condensed. 
In the many-body simulation, we observe that during the time evolution all 4 natural orbitals become significantly occupied, already after 4 cycles. 
We can therefore not claim that we reached numerical convergence 
with $M=4$ time-adaptive orbitals in this interaction regime, 
and we will therefore not physically interpret the underlying natural orbitals as we did in Fig.~\ref{fig:Fig5} for weak repulsion. 
To reach numerical convergence, one would need to allow for additional time-adaptive orbitals. 
However, already one additional orbital for the same number of particles would substantially increase the configuration space of size $\binom{N+M-1}{N}$ such that the computational effort would exceed the scope of this work. 
Nevertheless, we can draw some conclusions. 
Starting as above from the barrier located at the crossing point, 
the time evolution of $P_\text{OUT}(t)$ at the mean-field and many-body levels are depicted in the inset of Fig.~\ref{fig:Fig4}. 
At first, we see that much less particles are involved in the tunneling process compared to the case of $\Lambda=2$. 
Secondly, the periods of the oscillations for the mean-field and many-body descriptions are different. 
The equivalence of both approaches is broken already during the first cycle. 

Another interesting observation is that the choice of the initial condition becomes important for stronger repulsion, meaning that the dynamics of $P_\text{IN}(t)$ starting in the IN region and of $P_\text{OUT}(t)$ starting in the OUT region are no longer equivalent (not shown). It is worth mentioning that this is very different compared to the (one-dimensional) symmetric double-well system. 
In the latter, one would never observe any differences between $P_L(t)$, when starting in the left well, and $P_R(t)$, when starting in the right well, no matter how large the interaction strength is. This is due to the perfect spatial symmetry of the external potential which is not given in the 2D circular trap. 
Since the two IN and OUT subsystems have different topologies and sizes, 
one cannot expect their level structures to be equivalent; This leads to different dynamics for the initial conditions because stronger interaction means that more than just the lowest-in-energy time-adaptive orbitals are involved.

\subsection{Tunneling dynamics of vortex states}\label{dynvortex}

In the previous subsection, we have investigated the tunneling 
and fragmentation dynamics for bosons with zero total angular momentum, $L=0$. 
Here, we report on the tunneling dynamics of vortex states 
with definite total angular momentum $L=N$, and explore both similarities 
and differences compared to the dynamics of $L=0$. 
We will again discuss non-interacting and interacting bosons separately.

\subsubsection{The non-interacting system}

\begin{figure}[!]
\includegraphics[angle=-90,width=0.6\textwidth]{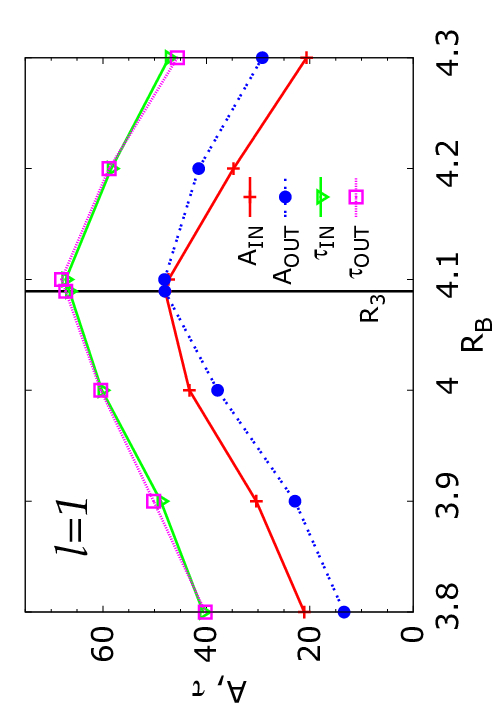}
\caption{(Color online) Dependence of the amplitude $A$ and period $\tau$ 
(plotted on the same axis) on $R_B$ in the non-interacting system, i.e., the single-particle case. 
The orbital angular momentum is $l=1$. 
The indices IN and OUT denote the two different initial conditions considered. 
Close to the crossing point $R_3=4.089$, 
the period and amplitude of the density oscillations are maximal and the two initial conditions give equivalent results for the tunneling dynamics. 
Leaving the crossing point, both the amplitude and period become smaller. 
See text for more details. 
All quantities are dimensionless.
}
\label{fig:Fig6}
\end{figure}

In this subsection, as a starting point, 
we investigate the dynamics of vortex states where the constituent particles do not interact, i.e., the single-particle case with 
orbital angular momentum $l=1$ [see Eq.~(\ref{eq_A1}) in Appendix \ref{AppA} for more details]. 
We first study the impact of the barrier location $R_B$ on the tunneling dynamics. 
We compare the periods and amplitudes of the density 
oscillations between the IN and OUT regions for several values of $R_B$ in Fig.~\ref{fig:Fig6}. 
We recall that for $l=1$ the crossing point in the non-interacting system is located at $R_3=4.089$, 
which is larger compared to the non-interacting system with $l=0$ where the crossing point is located at $R_1=3.271$ (see Fig.~\ref{fig:Fig2}). 
Now, for $l=1$ and barrier position $R_3$, 
both the amplitude $A$ and period $\tau$ are maximal with values of $A=48$ particles and $\tau=66.4$. 

In order to calculate the characteristic tunneling time $\tau_2$ of a vortex state, 
we prepared the two ground-state orbitals with $l=1$ for the IN region ($\phi^\text{IN}_0$) and for the OUT region ($\phi^\text{OUT}_0$, cp. Fig.~\ref{fig:Fig1}), respectively, and calculated $J_2=|\langle \phi^\text{OUT}_0|\hat{h}|\phi^\text{IN}_0 \rangle|=4.704\cdot 10^{-2}$. 
This in turn gives $\tau_2=\pi/J_2=66.786$ which is in a very good agreement with the result from Fig.~\ref{fig:Fig6}. 
Compared to the case of $l=0$, the vortex state tunnels almost twice as fast between the two parts of the trap. 
This can be anticipated from its higher energy and 
the lower effective barrier under which it has to tunnel through. 
We recall that the barrier height is kept fixed at $B=1$.
Moreover, the dynamics give similar results at $R_3$ for the two initial conditions considered. 
Our study of the interacting vortex states will therefore focus on the dynamics 
at the corresponding crossing points where the bosons are initially prepared in the OUT region.

\subsubsection{The interacting system}

We now investigate the impact of the 
interparticle repulsion $\Lambda$ on the tunneling dynamics of a vortex state with $L=N$. 
It turns out that at the many-body level, 
vortex states are more sensitive to the strength of the repulsion
than states with $L=0$ (see details below). 
In other words, many-body effects set in at weaker interactions. 
Thus, we split the discussion for the regime of weak repulsion ($\Lambda=0.2$) and strong repulsion ($\Lambda=2$). 
Note that the regimes for $L=N$ are different than for $L=0$, 
see the discussion in Sec.~\ref{Conclusion}. 

\begin{figure}[!]
\includegraphics[angle=-90,width=0.6\textwidth]{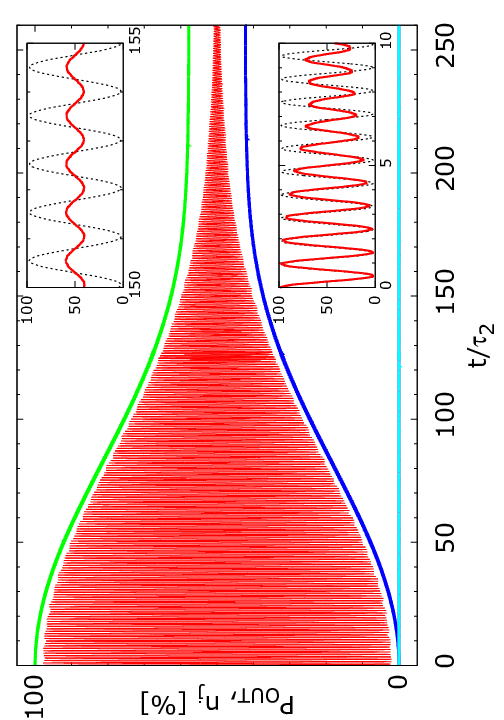}
\caption{(Color online) Many-body ($M=4$) tunneling dynamics of a vortex state at the crossing point $R_4=4.093$. 
The number of particles is $N=100$, the total angular momentum is $L=N$, 
and the interaction strength is $\Lambda=0.2$. 
The bosons are released from the OUT region. 
The damped oscillation of $P_\text{OUT}(t)$ (solid red) is enveloped 
by the natural occupations of $\alpha_1$ and $\alpha_2$ (solid green, blue). 
The remaining natural orbitals do not become significantly occupied (solid magenta, light blue; curves lie atop of each other). 
Upper inset: Dynamics of $P_\text{OUT}(t)$ for the GP, mean-field ($M=1$, dotted gray) and many-body ($M=4$, solid red) descriptions between $t=150\,\tau_1$ and $t=155\,\tau_1$. 
The system's parameters are the same as in the main figure. 
The mean-field description does not account for the density collapse; The vortex state is already fragmented. 
However, the tunneling frequencies of the GP and MCTDHB(4) results are essentially the same. 
Lower inset: Same simulation for strong repulsion, $\Lambda=2$, at the respective crossing point $R_B=4.123$. 
The other parameters of the system are the same as in the main figure. 
The dynamics obtained from the mean-field (dotted gray) and many-body descriptions (solid red) start to deviate from each other already after a few cycles, both in frequency and amplitude. 
After 10 cycles, the four natural orbitals are already macroscopically occupied (not shown). 
See text for more details. 
All quantities are dimensionless.
}
\label{fig:Fig7}
\end{figure}

For $\Lambda=0.2$, 
we put the barrier at $R_4=4.093$ (see Fig.~\ref{fig:Fig2}b) and prepare the bosons in the OUT region. At the mean-field level, the vortex state tunnels essentially unperturbed between the IN and OUT regions, i.e., there is no damping of the amplitude observable. 
The result is thus very similar to the one from the non-interacting 
case and is therefore not discussed further (see upper inset of Fig.~\ref{fig:Fig7}). 

The many-body dynamics on the contrary are more intriguing. 
Fig.~\ref{fig:Fig7} shows the MCTDHB(4) results for the time evolution of $P_\text{OUT}(t)$ together with the corresponding natural occupation numbers. 
The initially-coherent system $[n_2(0)=\mathcal{O}(10^ {-8})]$ evolves to a two-fold fragmented condensate 
where the two time-dependent natural orbitals 
both carry angular momentum $l=1$. 
The remaining two orbitals do essentially not participate 
in the dynamics since their occupations stay below 0.1\% throughout the time evolution. 

\begin{figure}[!]
\includegraphics[angle=0,width=0.6\textwidth]{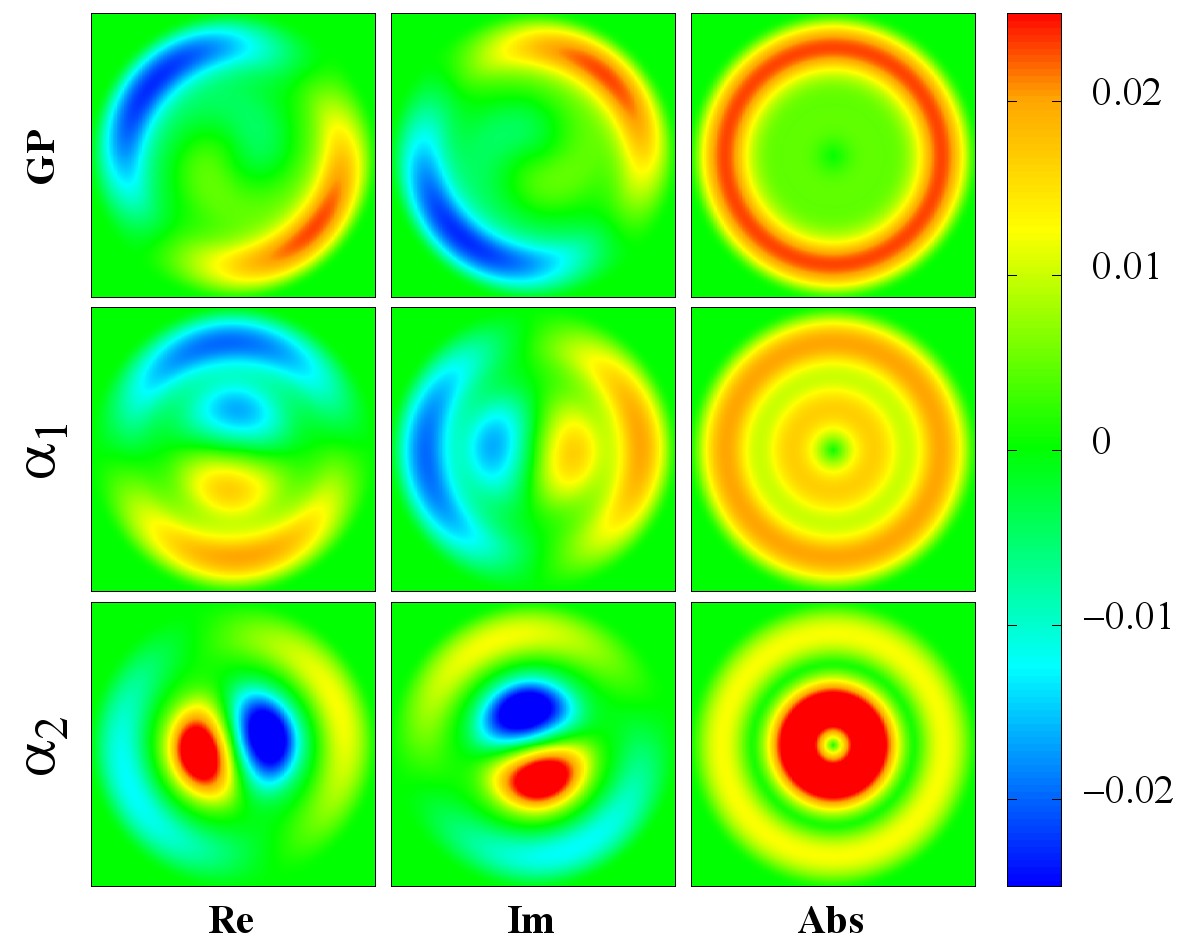}
\caption{(Color online) Real, imaginary, and absolute value of the GP, 
mean-field orbital (top panels) and of the first two natural orbitals $\alpha_1$ and $\alpha_2$ of the many-body simulation ($M=4$, middle and bottom panels) at $t=183.3\,\tau_2$. 
The system's parameters are the same as in the main figure of Fig.~\ref{fig:Fig7}. 
Whereas the mean-field orbital is localized in the external rim, 
the natural orbitals in the many-body computation are delocalized, covering both the IN and OUT regions. 
$\alpha_1$ only has a node in the trap's center, 
whereas $\alpha_2$ has a second, ring-shaped node at the barrier's position $R_4$.
See for comparison Fig.~\ref{fig:Fig5} and the text for more details. 
All quantities are dimensionless.
}
\label{fig:Fig8}
\end{figure}

The evolution of $P_\text{OUT}(t)$ shows damping of the oscillations. 
The damping resembles the case of $L=0$ where we observed the same phenomenon. 
The behavior at longer times is such that the system tends to distribute the 
particles equally between the IN and OUT regions, and the density oscillations get suppressed. 
The GP description does not predict the density collapse 
of a vortex state and thus a many-body treatment of the dynamics is necessary. 
This conclusion is supported by Fig.~\ref{fig:Fig8} 
which shows the structure of the first two natural orbitals from the many-body 
computation as well as the corresponding mean-field orbital at an intermediate time step $t\approx 183.3\,\tau_2$. 
The GP theory does not account for the occupation of the trap center and only predicts particles in the external rim. 
In contrast to that, the delocalized orbitals of the many-body computation 
show significant occupations of both the IN and OUT regions. 
In the trap's center, 
the characteristic nodal structure of vortex states can be observed. 
The second orbital $\alpha_2$ has an additional ring-shaped node, 
matching the location of the barrier at $R_4$.

We repeated the same simulation for strong repulsion $\Lambda=2$ at the respective crossing point $R_B=4.123$. 
By observing that already for this interaction strength all 4 natural orbitals become macroscopically occupied, we deduce that for this value of $\Lambda$ one has already left the weak repulsion regime for vortex states. 
Thus, vortex states are much more sensitive to interparticle repulsion in comparison to states with $L=0$. We again stress that it is necessary to include more time-adaptive orbitals to reach numerical convergence. 
Nevertheless, already $M=4$ orbitals are enough to expose substantial differences 
between the mean-field and many-body predictions already after a few cycles, both concerning the amplitude and frequency of the density oscillations (see lower inset of Fig.~\ref{fig:Fig7}; note the different time scales).

\section{Concluding remarks}\label{Conclusion}

In conclusion, in the present work we have studied static properties 
and particularly the out-of-equilibrium tunneling dynamics of BECs and vortex states 
in a 2D radial double well. 

On the statics side, we showed that angular momentum and repulsion between the bosons affect the location of the ground state, i.e., whether it is energetically favorable for the bosons to occupy either the IN or the OUT region of the trap. 
For a certain critical radius, which we termed the crossing point, the ground-state energies of the IN and OUT subsystems are equivalent. 
The position of the crossing point depends on both angular momentum and repulsion strength. 
The impact of angular momentum is however significantly stronger, i.e., it shifts the crossing point to much larger radii. 

On the dynamics side, we observed several similarities as well as clear differences 
between the tunneling of BECs with $L=0$ and of vortex states with $L=N$. 
At first, we demonstrated periodic, Rabi-like tunneling between the IN and OUT regions 
in the non-interacting and weakly-interacting regimes for both states with $L=0$ and vortex states with $L=N$. 
The density oscillations are most pronounced when the barrier is located at the corresponding crossing points. 
Leaving the crossing points leads to density oscillations 
with shorter periods and smaller amplitudes, i.e., less particles are involved in the tunneling process. 
However, by comparing the characteristic tunneling times, 
we found that the tunneling of a vortex state with $L=N$, which is higher in energy, 
takes place on a (much) shorter time scale, 
being almost twice as fast as for states with $L=0$. 

For both values of angular momentum considered, the time-dependent GP 
equation fails to describe the long-time tunneling dynamics as soon as the particles interact. 
For weak repulsion, both systems with $L=0$ and $L=N$ become essentially two-fold fragmented.

The development of fragmentation is accompanied by damping of the density oscillations between the IN and OUT regions. 
The stronger the interaction $\Lambda$ is, the faster the density oscillations are suppressed. 
The regime of weak repulsion for $L=0$ is apparently more extended, 
which in turn means that vortex states are more sensitive to repulsion. 
We deduce this from the fact that already for $\Lambda=2$
(which corresponds to weak repulsion for states with $L=0$) 
more than two natural orbitals 
are significantly occupied 
in the many-body dynamics of vortex states with $L=N$ 
at long simulation times.
We have additionally shown that for the strongly repulsive systems 
($\Lambda=6$ for $L=0$ and $\Lambda=2$ for $L=N$), 
at least 4 time-adaptive interaction-dressed orbitals in the MCTDHB theory 
are necessary in order to describe the many-body tunneling dynamics faithfully. 
In both cases, the many-body dynamics quickly start to deviate 
from the corresponding GP predictions.

The present work shows that the tunneling dynamics of BECs and of vortex states in a 2D radial double well 
is many-body in nature.
Furthermore, many-body effects set in at even weaker interactions when the tunneling system carries angular momentum.
Obviously, with increasing interaction more and more many-body excited states are involved,
where both radial and angular excitations can combine to assemble states of definite total angular momentum.  
These may allow for an even richer and more intricate out-of-equilibrium dynamics than reported here. 

\section*{acknowledgments}

Computation time on the Cray XE6 cluster Hermit and the Cray XC40 cluster Hornet 
at the High Performance Computing Center Stuttgart (HLRS), as well as 
financial support of the Heidelberg Graduate School of Fundamental 
Physics (HGSFP) and the Deutsche Forschungsgemeinschaft (DFG) are gratefully acknowledged.

\appendix

\section{Further details on vortex states and total angular momentum}\label{AppA}

In this appendix, we give additional details on the numerical preparation of vortex states with $L=N$, 
as well as on the conservation of the angular momentum during
the propagation in time of the many-body wave-functions.

The numerical preparation of a vortex state in the 2D radial double-well potential
can be achieved by multiplying radially symmetric, real-valued functions $f_j(r,t),\,1 \leq j \leq M$, with a phase of integer value:
\beq\label{eq_A1}
 \phi_j(\vec{r},t)=f_j(r)\,e^{il_j\theta}, \quad\quad 1 \leq j \leq M,
\eeq
where $\theta$ is the phase and $l_j$ the integer angular momentum per particle. 
The orbitals $\{\phi_j(\vec{r},t)\}$ are then propagated in imaginary time, leading to the initial vortex states from which we study the dynamics in the main text. In principle, the individual orbitals involved can have different values $l_j$. 
In this work, we start from essentially condensed systems and we thus concentrate 
on states where the underlying orbitals have the same $l_j=l=1$.

During the time evolution, we measure the angular momentum of the underlying natural orbitals via 
\beq
 l_k=\langle \alpha_k(t)|\hat{L}_z|\alpha_k(t)\rangle, \quad\quad  1 \leq k \leq M.
\eeq
The angular momentum operator in z-direction is defined as $\hat{L}_z=-i\,\left(x\frac{\partial}{\partial y}-y\frac{\partial}{\partial x}\right)$. For the regime of weak repulsion, 
we observe that $l_k=1$ for the macroscopically-occupied orbitals is nicely conserved throughout the whole time evolution.
The total angular momentum $L$ is measured in our computations via the quantity
\beq
 L=\sum_{j,k}(L_z)_{jk}(t)\,\rho_{jk}(t),
\eeq
with $(L_z)_{jk}(t)=\langle \phi_j(t)|\hat{L}_z|\phi_k(t)\rangle$. $L$ serves as a good quantum number for the dynamics of both values of angular momentum considered in the main text.

Our study suggests the implementation of a projection operator onto
good-total-angular-momentum-eigenstates of the full many-body Hamiltonian Eq.~(\ref{Hamiltonian}) as a useful numerical development, extending our current method of preparing vortex states. 
In such a way, one would be able to prepare the system in many-body eigenstates of, e.g., non-integer total angular momentum per particle, $L/N$. 
This would possibly allow for a more sophisticated study of vortex states' dynamics, in particular for stronger interactions, as well as the study of the 
evolution in time of the spatially-partitioned 
many-body vortices proposed in \cite{klaiman2014spatially}.

\section{Numerical convergence for $L=0$}\label{AppB}

\begin{figure}[!]
\includegraphics[angle=-90,width=0.6\textwidth]{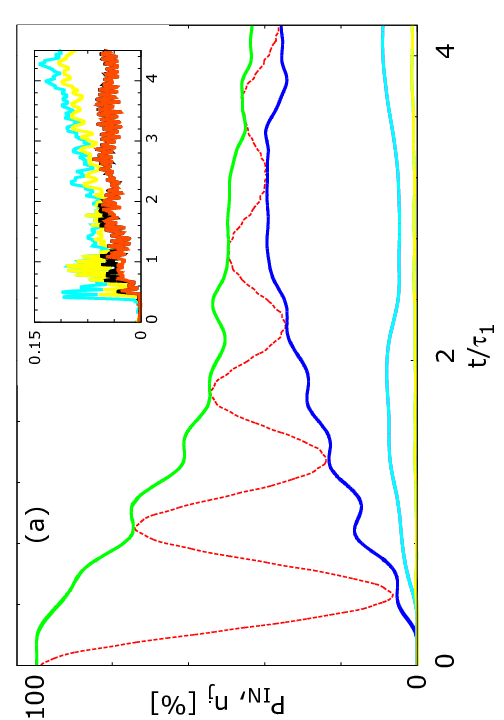}
\includegraphics[angle=-90,width=0.6\textwidth]{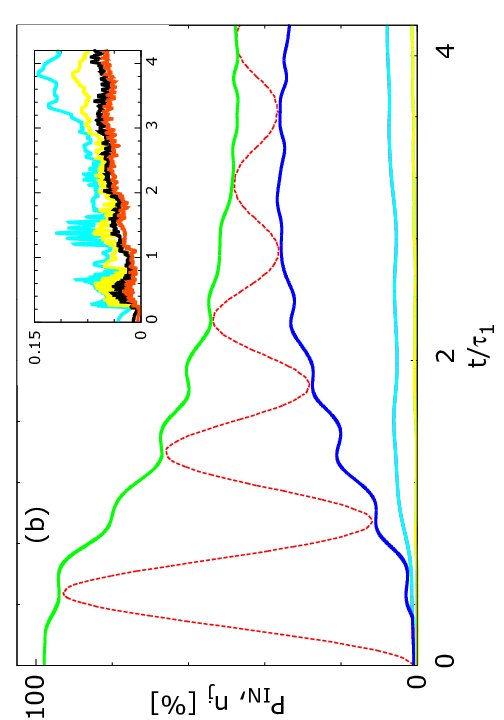}
\caption{(Color online) (a): Many-body tunneling dynamics for $N=10$ particles. 
The number of orbitals is $M=10$, the angular momentum is $L=0$, and the interaction strength is $\Lambda=2$. 
The bosons are released from the IN region. 
The dashed red curve denotes the time evolution of $P_\text{IN}(t)$. 
Only the first 4 natural orbitals become macroscopically occupied 
[solid curves from top to bottom: $n_1(t)$ in green, $n_2(t)$ in blue; $n_3(t)$ in magenta, 
and $n_4(t)$ in light blue (atop of the $n_3(t)$ curve)]. 
The $n_5(t)$ and $n_6(t)$ natural occupations (in light green and yellow; bottommost solid curves lying atop each other) 
carry only 1\% of the particles. Inset: The remaining natural occupation numbers 
[solid curves from top to bottom: $n_7(t)$ in light blue, $n_8(t)$ in yellow; $n_9(t)$ in black, 
and $n_{10}(t)$ in brown], all staying below 0.15\%. 
The fact that only the first 4 natural orbitals are significantly occupied justifies the description of the tunneling dynamics with $M=4$ time-adaptive orbitals for $\Lambda=2$. 
(b) The same simulation as in (a) but releasing the particles from the OUT region,
which gives similar but not identical results. 
Most important, only the first 4 natural orbitals become macroscopically occupied. 
All quantities are dimensionless.
}
\label{fig:FigA1}
\end{figure}

The purpose of this appendix is to justify the restriction to $M=4$ 
time-adaptive orbitals within the MCTDHB(M) theory in case of weak repulsion for $L=0$. 

Fig.~\ref{fig:FigA1} shows the tunneling dynamics for $N=10$ particles, allowing for $M=10$ time-adaptive orbitals. 
The interaction parameter has been set to $\Lambda=\lambda_0(N-1)=2$ as in the main text (see Fig.~\ref{fig:Fig4}). 
The tunneling dynamics are again dominated by two natural orbitals which carry 
the majority of the occupation probability. 
In addition to that, there are two more natural orbitals, occupied by roughly 10\% of the bosons. 
However, the most important observation is that the remaining 6 natural orbitals 
do not become significantly occupied, i.e., $n_{j>4}(t)\leq0.15$\%. 
This proves that numerical convergence of the many-body tunneling 
dynamics in the 2D radial double well with $M=4$ time-adaptive orbitals is reached.


\begin{thebibliography}{99}
\bibitem{realunits} To convert to real units, we choose as an example a condensate of $^{87}$Rb atoms. Choosing $d=2\,\mu m$, the crater radius in our simulations is given by $R_C=18\,\mu m$. As a unit of energy we obtain $29.1$ Hz which coincides with the barrier height $B$. The height of the crater wall is $C=5.82$ kHz.
\end{thebibliography}
\end{document}